\documentclass[portugues]{sobraep}
\usepackage{hyperref}
\usepackage[sort]{natbib}
\usepackage{amsmath}
\usepackage{siunitx}
\DeclareUnicodeCharacter{2212}{-}

\titulo{Uncertainty-Aware Neural Networks for Fuzzy Dark Matter Model Selection from  \texorpdfstring{$  x_{\rm HI}$}{x_HI} Measurements}

\title{Uncertainty-Aware Neural Networks for Fuzzy Dark Matter Model Selection from  \texorpdfstring{$  x_{\rm HI}$}{x_HI} Measurements}

\author{Bahareh Soleimanpour Salmasi$^{1}$, S. Mobina Hosseini$^{1}$ \\
	\normalsize $^{1}$Department of Physics, Shahid Beheshti University, P.O. Box 19839-69411, Tehran, Iran\\
	\normalsize e-mail: physics.bahare@gmail.com, mobinahosseini954@gmail.com}

\begin{document}
\DeclareSIUnit\angstrom{\text {Å}}
\maketitle

\begin{abstract}
	The nature of dark matter remains a central question in cosmology, with fuzzy dark matter (FDM) models offering a compelling alternative to the cold dark matter (CDM) paradigm. We explore FDM scenarios by performing 21-cm simulations across a parameter space with $f_{\mathrm{FDM}} \in [0.02, 0.10]$ and $m_{\mathrm{FDM}} \in [10^{-24}, 10^{-21}]\,\mathrm{eV}$, obtaining global neutral hydrogen fractions ($x_{\mathrm{HI}}$) for each model. Observational $x_{\mathrm{HI}}$ data and associated uncertainties from JWST are incorporated by estimating full probability density functions (PDFs) for both $x_{\mathrm{HI}}$ and redshift $z$ using Bayesian inference with the No-U-Turn Sampler (NUTS), yielding non-Gaussian multivariate uncertainty distributions. A hybrid machine learning framework is then trained on these observational PDFs to learn both central values and correlated uncertainties in $x_{\mathrm{HI}}$ and $z$, iteratively refining its parameters in each training epoch through direct incorporation of the multivariate PDFs derived from observational constraints. We then compare the simulation outputs to the machine-learned observational trends to identify the most consistent models. Our results indicate that FDM models with $m_{\mathrm{FDM}} \simeq 10^{-22}\,\mathrm{eV}$ and $f_{\mathrm{FDM}} \simeq 0.04$ best match current data, while lighter masses are strongly constrained. By integrating simulations and machine learning in an uncertainty‑aware framework, this work explores the physics of the early Universe and guides future studies of 21‑cm cosmology and reionization.
\end{abstract}


\section{Introduction}\label{SEC1}

Understanding how and when the intergalactic medium (IGM) transitioned from neutral to ionized is traced by the evolution of $x_{\rm HI}$. Galaxies at high redshift exhibit a distinct ultraviolet (UV) spectrum with a sharp drop at wavelengths shorter than 912\,\si{\angstrom} (rest-frame) \citep{wilkins2011ultraviolet}, caused by absorption from neutral hydrogen in stellar atmospheres, the interstellar medium, and the IGM. At $z \geq 6$, the presence of neutral hydrogen in the IGM leads to almost complete absorption of light shortward of the Lyman-$\alpha$ (Ly$\alpha$) transition at 1216\,\si{\angstrom}, due to resonant scattering \citep{zavala2023dusty,steidel2003lyman}.

As the Universe’s first stars and galaxies formed, their UV radiation carved out expanding ionized bubbles \citep{furlanetto2005taxing}, progressively reducing $x_{\rm HI}$ from nearly unity to zero \citep{holzer1977neutral}. The timing and pace of this reionization process encode both astrophysical conditions and the influence of dark matter on small scales \citep{desjacques2018impact,morales2010reionization}. Observational probes, including Gunn--Peterson troughs \citep{fan2006observational}, Ly$\alpha$ damping wings \citep{hayes2015lyman}, Ly$\alpha$ emitter clustering \citep{hutter2015clustering}, and galaxy luminosity functions \citep{salpeter1986galaxy}, provide constraints on the ionized fraction and the sources driving reionization.

Because the reionization history is sensitive to the abundance of small-scale structure, it can serve as a probe of non-standard dark matter models. FDM scenarios posit ultra-light axions whose wave-like quantum pressure suppresses density fluctuations on sub-kiloparsec scales \citep{hui2017ultralight,marsh2016axion,hu2000fuzzy}. This suppression delays low-mass halo collapse \citep{kulkarni2022halo}, postpones star formation \citep{mayer2022new}, and extends the reionization timeline relative to CDM predictions \citep{burigana2022cosmic}. Such effects imprint distinctive signatures on $x_{\rm HI}$ evolution \citep{shen2024thesan}, including a deeper 21-cm absorption trough at lower redshift \citep{lidz2018implications} and a slower decline in the neutral fraction at $z \gtrsim 8$ \citep{lazare2024constraints,dome2024modelling}.

Recent work has leveraged $x_{\rm HI}$ evolution to constrain both IGM models and early structure formation. \citet{hosseini2023new} used $x_{\rm HI}$ measurements to rule out inconsistent thermal histories within the semi-numerical framework \texttt{21cmFAST}. \citet{flitter2022closing} forecasted that the Hydrogen Epoch of Reionization Array (HERA) \citep{abdurashidova2022first,abdurashidova2022hera} could detect FDM fractions as low as 1\% in the mass range $10^{-25}\,\mathrm{eV} \lesssim m_{\rm FDM} \lesssim 10^{-23}\,\mathrm{eV}$, improving existing bounds by up to an order of magnitude. Extending this, \citet{lazare2024constraints} combined ultraviolet luminosity functions, high-redshift quasar spectroscopy, and 21-cm power spectrum limits to constrain the FDM fraction to below 16\% for $m_{\rm FDM} = 10^{-23}\,\mathrm{eV}$, and down to 1\% for $10^{-26}\,\mathrm{eV}$, using a machine learning--based inference pipeline to explore high-dimensional cosmological parameter spaces.

Additional studies highlight the impact of FDM on the global 21-cm signal. \citet{sarkar2022exploring} found that FDM-induced delays in structure formation alter both the timing and amplitude of the signal, with HERA potentially sensitive to particle masses up to $10^{-19}$--$10^{-18}\,\mathrm{eV}$. The tentative EDGES detection of a pre-reionization 21-cm absorption feature \citep{monsalve2018results}---if confirmed---implies early Ly$\alpha$ coupling at $z \simeq 20$, placing a lower bound of $m_{\rm FDM} \gtrsim 5 \times 10^{-21}\,\mathrm{eV}$ to ensure timely star formation in low-mass halos \citep{lidz2018implications}. Post-reionization, FDM cosmologies predict enhanced bias and filamentary morphology in neutral hydrogen distributions \citep{dome2024modelling}, with conditional normalizing flows offering a promising framework to capture the non-linear mapping between axion mass and the spatial structure of $x_{\rm HI}$.

Convolutional neural networks (CNNs) are effective at extracting and reconstructing spatial features while maintaining low computational cost, offering complementary strengths for 21‑cm signal analysis \citep{purwono2022understanding,cong2014minimizing}. In contrast, recurrent neural networks (RNNs) are well suited to time‑series analysis because of their ability to capture patterns within sequential data \citep{grossberg2013recurrent}, making them particularly useful for studying the inherently time‑dependent signals from the EoR \citep{faaique2024overview}. Integrating simulations with these neural network architectures helps break parameter degeneracies, such as distinguishing between X‑ray heating and Ly$\alpha$ coupling \citep{kacprzak2022deeplss}.

The thermal history of the IGM and the 21-cm signal is being probed by radio facilities such as the LOw-Frequency ARray (LOFAR) \citep{greig2021interpreting,ghara2020constraining}, the Murchison Widefield Array (MWA) \citep{tingay2013murchison}, the Giant Metrewave Radio Telescope (GMRT) \citep{chowdhury2022giant,paciga2011gmrt}, and the forthcoming Square Kilometre Array (SKA) \citep{weltman2020fundamental}. Meanwhile, James Webb Space Telescope (JWST) observations \citep{gardner2023james,umeda2024jwst} suggest higher $x_{\rm HI}$ at $z > 8$ than many CDM models predict \citep{umeda2024jwst,jin2023nearly,morales2021evolution}. These emerging tensions raise a key question: can FDM reconcile theoretical expectations with these observations, or are alternative astrophysical mechanisms required?

To address this, we present a hybrid framework combining \texttt{21cmFirstCLASS} simulations \citep{flitter202321cmfirstclass} with uncertainty-aware neural networks. \texttt{21cmFirstCLASS} incorporates FDM scenarios in which the wave-like nature of ultra-light axions suppresses low-mass halo formation, reshaping the spatial and temporal distribution of early ionizing sources. We compare simulated $x_{\rm HI}$ predictions against constraints from Ly$\alpha$ forest transmission, damping wing absorption in high-redshift quasars and gamma-ray bursts, and post-reionization HI bias, in FDM cosmology \citep{barkana2001beginning,zaroubi2012epoch,choudhury2022short}. By fine-tuning on $x_{\rm HI}$ measurements and explicitly modeling non-Gaussian uncertainties, we extract posterior constraints on the underlying dark matter model.

In Section~\ref{SEC2}, we review the main equations governing FDM, 21-cm cosmology, and galaxy properties, and outline observational techniques for HI. Section~\ref{SEC3} details the simulations and the parameters employed, and Section~\ref{SEC4} outlines the evaluation procedures and performance metrics. Sections~\ref{SEC5} and \ref{SEC6} then present the results and conclusions, respectively.

\section{Backgrounds and Data}\label{SEC2}

In this section, we review the theoretical and observational foundations relevant to our analysis. Section~\ref{SEC2.1} outlines the key properties of FDM and its impact on structure formation. Section~\ref{SEC2.2} summarizes the basics of 21-cm cosmology and its sensitivity to reionization history. 
Section~\ref{SEC2.3} describes the observational datasets used to constrain $x_{\rm HI}$.

\subsection{Fuzzy Dark Matter}\label{SEC2.1}

A well‑known feature of the $\Lambda$CDM model of cosmic structure formation \citep{white1991galaxy} is that dark matter halos in N‑body simulations exhibit nearly universal spherically averaged density profiles, well described by the Navarro–Frenk–White (NFW) form \citep{navarro1996structure} 
\begin{equation}
    \frac{\rho(r)}{\rho_{crit}}=\frac{\delta_c}{(r/r_s)(1 + r/r_s)^2},
    \label{eq1}
\end{equation}
 \noindent $\delta_c$ is the characteristic overdensity of the halo, $r = r_s$ is the halo’s scale radius, and $  \rho_{crit} = 3H^2/8\pi G$ is the critical density \citep{navarro1997universal}. The density diverges as $\rho \simeq r^{-1}$, which is referred to as the central cusp, for $r \ll r_s$ \citep{burkert2020fuzzy}. 

Observations often show that rotation curves of galaxies, especially those with low masses dominated by dark matter, have flat distributions of inner dark matter density, which goes against this fundamental feature of CDM halos. According to several sources \citep{moore1994evidence,burkert1995structure,gentile2004cored,genzel2020rotation}, so-called dark matter cores have a flat density profile within a core radius ($r_0$) and a finite central density ($\rho_0$) configuration of self-gravitating isothermal spheres \citep{weinberg2015cold,li2020comprehensive}.

The origin of the CDM core–cusp problem is extensively contested \citep{governato2012cuspy,ogiya2014core}.
A particular class of models posits that violent fluctuations in the gravitational potential within the inner regions of galaxies are induced by baryonic processes, such as perturbations from the clumpy, turbulent interstellar medium or powerful galactic winds that release a significant portion of an initially gravitationally dominant gas component \citep{chan2015impact,benitez2019baryon,freundlich2020model,faucher2023key}.

A potentially more intriguing alternative is that the core–cusp problem indicates concealed characteristics of the dark matter particle that are not considered in conventional cosmological simulations. FDM \citep{hu2000fuzzy} is a notable situation that has garnered significant attention recently. The quantum-mechanical wave-like behavior of ultra-light particles (e.g., axions) on the galactic scale is reflected in the term "fuzzy". Their great de Broglie wavelength ($  \lambda_{dB}\simeq \hbar/m\nu$) produces a "fuzzy" density distribution, therefore preventing small-scale structure growth. The detected dark matter cores would consequently be soliton cores arising from an equilibrium between quantum pressure, as specified by the uncertainty principle, and gravitational forces, with the observed core directly correlating to the mass $  m$ of the FDM particles. The fundamental properties are defined by resolving the coupled Schr\"{o}dinger–Poisson equation \citep{chiu2025boosting}.

The primary self-consistent cosmological 3D simulation of FDM halo generation is introduced by \citet{schive2014understanding}. They verified that all halos that form a unique, gravitationally self-bound solitonic core exhibit a uniform core density distribution.
For radii $  r \leq 3 r_0$, it can be accurately represented by the empirical relation.
The soliton density profile is approximated by \citep{schive2014cosmic}

\begin{eqnarray}
  \rho(r) &\simeq& 1.9 \left(\frac{m}{10^{-23}\,\mathrm{eV}}\right)^{-2} \nonumber \\
&&  \times \left(\frac{(r_0/\rm kpc)^{-1}}{\left(1 + 9.1 \times 10^{-2}\,(r/r_0)^2\right)^2}\right)^4 \frac{M_\odot}{\rm pc^3},
\label{eq:2}
\end{eqnarray}

\noindent where the core radius $  r_0$ scales with halo virial mass and redshift. 

\citet{schutz2020subhalo} analyzed two of the faintest, intensely dark matter-dominated Milky Way dwarf galaxies and determined that the FDM particle mass m is between $3.7-5.6 \times10^{-22} \mathrm{eV}$. This outcome, however, is predicated on the supposition that the stellar component is entirely embedded within the soliton core and that these diffuse satellite galaxies, which are located at galactocentric distances of $20 \rm kpc$ \citep{laevens2015new} and $26 \rm kpc$ \citep{laevens2015sagittarius}, are not strongly tidally perturbed and are in virial equilibrium. \citet{safarzadeh2020ultra} analyzed two more distant Milky Way dwarf spheroidals, with galactocentric distances of $88 \rm kpc$ and $138 \rm kpc$, respectively \citep{kormendy2016scaling}.
They inferred axion masses of the order of $10^{-21} \mathrm{eV}$, with a degree of dependence on the unknown halo virial mass. \citet{wasserman2019spatially} examined the ultra-diffuse, intensely dark matter-dominated galaxy Dragonfly 44 and determined soliton masses of approximately $3 \times 10^{-22} \mathrm{eV}$. \citet{li2020testing} demonstrated that the origin of the central molecular zone dynamics in the Milky Way would be better explained by a soliton core, which corresponds to a boson mass of approximately $2-7 \times 10^{-22} \mathrm{eV}$. This matter necessitates a dense, compact central mass concentration. Lastly, \citet{davies2020fuzzy} demonstrated that the density profile of soliton cores could be influenced by supermassive black holes located in the centers of galaxies, thereby limiting the FDM particle mass. Hence, these investigations suggest that the FDM particle mass is within $  10^{-27} \mathrm{eV} \leq m \leq 10^{-20}\rm eV$. The quantum pressure applied by a dark matter fluid composed of minuscule particles suppresses small-scale fluctuations \citep{marsh2016axion,hui2017ultralight}. As the mass decreases, the suppression strengthens.

The FDM model possesses theoretical merit, as it addresses several challenges that undermine the CDM model when its simulations are compared to observations, including the core-cusp problem, the missing satellites problem, and the too-big-to-fail problem \citep{hu2000fuzzy,bullock2017small}, despite the existence of alternative explanations such as selection effects and baryonic feedback. Furthermore, current research indicates that particular values of the FDM mass and fraction could mitigate the tensions associated with $H_0$ \citep{blum2021gravitational} and $  S_8$ \citep{allali2021dark,lague2022constraining,ye2023alleviating}. FDM can provide an alternate explanation to the Stochastic Gravitational Wave Background (SGWB) regarding the NANOGrav \citep{agazie2023nanograv,afzal2023nanograv} observation of pulsar time correlations \citep{chowdhury2024ultralight}. The FDM trait of suppressing small-scale fluctuations influences structure development, prompting several investigations to utilize cosmological and astrophysical observables to constrain the parameter space of the FDM mass $  m_{\rm FDM}$ and fraction $  f_{\rm FDM}$. Observable phenomena encompass CMB, large-scale structure (LSS) observations derived from Planck \citep{hlozek2015search,lague2022constraining}, the Dark Energy Survey (DES) \citep{baxter2019dark}, the Ly$\alpha$-forest \citep{irvsivc2017first,rogers2021strong}, the high-redshift UV luminosity functions, and the CMB optical depth to reionization \citep{bozek2015galaxy,corasaniti2017constraints,winch2024high}.

The FDM is also referred to as Wave Dark Matter ($\psi DM$) \citep{schive2014understanding} because of its quantum wave-like features dictated by the Schrödinger-Poisson equations, Ultra-Light Dark Matter (ULDM) \citep{hui2017ultralight} indicating its particle mass, and Bose-Einstein condensate dark matter \citep{sikivie2009bose} due to its coherent bosonic state at cosmic scales. It is known as Scalar Field Dark Matter (SFDM) \citep{matos2000quintessence} when represented as a classical scalar field, Quantum Dark Matter \citep{helfer2018quantum} to highlight its non-classical pressure effects, and Axion-like Dark Matter \citep{marsh2016axion} due to its proposed association with axion-like particles.

\subsection{21-cm Theory}\label{SEC2.2}
 
Radio telescopes are applied to quantify the intensity distribution of the sky, specifically, its Fourier transform. This measurement is conventionally represented in brightness temperature  $  T_{21}$, within the Rayleigh-Jeans regime. The brightness temperature of HI can be expressed in the following \citep{madau199721}
\begin{equation}
\begin{aligned}
  \delta T_{21}= &  \frac{T_S-T_{\rm CMB}}{1+z}\left(1-e^{-\tau_\nu}\right) \\
\approx &  \frac{T_S-T_{\rm CMB}}{1+z} \tau \\
\approx &  27 x_{\rm HI}\left(1+\delta\right)\left(\frac{\Omega_b h^2}{0.023}\right)\left(\frac{0.15}{\Omega_m h^2} \frac{1+z}{10}\right)^{1 / 2} \\
&  \times\left(\frac{T_S-T_{\rm CMB}}{T_S}\right)\left[\frac{\partial_r v_r}{(1+z) H(z)}\right] \rm mK,
\end{aligned}
\label{t_b}
\end{equation}
\noindent where $x_{\rm HI}$ is the neutral hydrogen fraction, $\delta$ is the spatial density fluctuations, $z$ is the redshift, $  T_s$ is the spin temperature, $T_{\rm CMB}$ is the CMB temperature,
 $\Omega_{\mathrm{b}}$ is the baryon density, $\Omega_{\mathrm{m}}$ is the matter density, $  h$ is the Hubble constant, $H(\it z)$ is the Hubble parameter, and $\partial_r v_r $ is velocity gradient along the line of sight, it is due to peculiar motions and bulk-flows of the gas. 

 The optical depth of a hydrogen cloud is \citep{pritchard201221} 
\begin{equation}
  \tau_\nu = \int \it d  s \left[ 1 - \exp \left( -\frac{E_{10}}{k_B T_S} \right) \right] \sigma_{0} \phi(\nu) n_{0},
\label{eq:tau_integral}
\end{equation}

\noindent where $  n_{0} = n_{\rm H}/4$  and $  n_H$ is the hydrogen density. $\sigma(\nu) = \sigma_{0} \phi(\nu)$ is the 21 cm cross-section and
$  \sigma_{0} = \frac{3c^{2} A_{10}}{8\pi \nu^{2}}$, where $A_{10} = 2.85 \times 10^{-15}~s^{-1}$ is the Einstein coefficient which is the spontaneous decay rate of the spin-flip transition \citep{pritchard201221}, and $  k_B$ is the Boltzmann constant. To ensure consistency and comparability, the line profile is normalized $\int \phi(\nu) {\rm d}\nu=1$, and the relation between frequency, $\nu$ and path, $  s$ can be evaluated as $  v = \left( \frac{\it{\rm d}v}{\it{\rm d}s} \right) s$ \citep{sobolev1957diffusion}.

The spin temperature signifies the thermal balance between various spin states of particles. When hydrogen atoms interact with radiation at the 21-cm wavelength, their spin states shift. The spin temperature reflects the average energy level of these spins and directly influences the observed intensity of 21-cm emission from the HI regions. Its formula is determined as follows

\begin{equation}
   T_S^{-1}=\frac{T_{\rm CMB}^{-1}+x_{\alpha}T_C^{-1}+x_{c}T_K^{-1}}{1+x_{c}+x_{\alpha}},
 \label{e6}
 \end{equation}
 where $T_C$ represents the color temperature of the Ly$\alpha$ radiation field at the Ly$\alpha$ frequency, $T_K$ is the gas kinetic temperature, and $x_c$ and $x_{\alpha}$ denote the coupling coefficients.
 
The equation itself is an expression for the inverse of the spin temperature, which is influenced by three main processes: the background radiation $T_{\rm CMB}^{-1}$, the WF effect $x_{\alpha}T_C^{-1}$, and collisions $x_cT_K^{-1}$. The denominator normalizes the equation, accounting for the relative strength of these processes in coupling the spin temperature to the kinetic temperature of the gas and the radiation field \citep{field1959spin,furlanetto2006cosmology}.

The Saha equation describes the balance between ionization and recombination processes in a plasma. It relates the ionization fraction to temperature and density.
For hydrogen, the Saha equation can be written as \citep{doughty2019evolution} 
\begin{equation}
  \frac{n_{\rm HII}}{n_{\rm H}} = \frac{1}{n_{\rm e}} \left( \frac{2\pi m_{\rm e} kT}{h^2} \right)^{3/2} e^{-E_{ion}/kT}, 
\end{equation}
\noindent where $  n_{\rm HII}$ is the number density of ionized hydrogen, $  n_{\rm H}$ is the total number density of hydrogen (neutral + ionized), $  n_{\rm e}$ is the number density of free electrons, $  E_{ion}$ is the ionization energy of hydrogen.

The evolution of the hydrogen ionized fraction $  x_i$ is expressed as \citep{bera2023studying}
\begin{equation}
\begin{aligned}
&  \frac{\it{\rm d}x_i}{\it{\rm d} z}=\left[C_p\left(\beta_i\left(T_{\rm CMB}\right)\left(1-x_i\right) e^{-\frac{h_{\rm p} v_\alpha}{k_{B} T_K}}-\alpha_i\left(T_K\right) x_i^2 n_{\rm H}(\it z  )\right)\right. \\
&  \left.\quad+\gamma_i\left(T_K\right) n_{\rm H}(\it z)\left(1-x_i\right) x_i+\frac{\dot{N}_\gamma}{n_{\rm H}(\it z)}\right] \frac{\it{\rm d} t}{\it{d} z},
\end{aligned}
\end{equation}
 \noindent where $  h_{\rm p} v_\alpha=10.2 \rm eV$, $  n_{\rm H}(\it z  )$ represents the appropriate number density of hydrogen atoms, and $  \dot{N}_\gamma$ is the rate of UV photons escaping into the IGM \citep{barkana2001beginning}. In this context, changes in the ionization fraction occur due to processes such as photoionization by CMB photons, recombination, collisional ionization, and photoionization by UV photons.
 
The photoionization coefficient is \citep{seager2000exactly,seager1999new}
\begin{equation}
  \beta_i\left(T_{\rm CMB}\right)=\alpha_i\left(T_{\rm CMB}\right)\left(\frac{2 \pi m_{\rm e} k_B T_{\rm CMB}}{h_{\rm p}^2}\right)^{3 / 2} e^{-E_2 s / k_B T_{\rm CMB}}.
\end{equation}

 The recombination coefficient is \citep{bera2023studying}
 \begin{equation}
   \alpha_i\left(T_K\right)=F \times 10^{-19}\left(\frac{a t^b}{1+c t^d}\right) m^3 s^{-1},
 \end{equation}
 where $  a=4.309$, $  b=-0.6166$, $  c=0.6703$, $  d=0.53$, $  F=1.14$ (correction factor) and $  t=T_K /\left(10^4 K\right)$.
 The Peebles factor is
 \begin{equation}
  C_p=\frac{1+K \Lambda(1-x) n_{\rm H}}{1+K\left(\Lambda+\beta_i\right)(1-x) n_{\rm H}},
\end{equation}
\noindent where $\Lambda=8.3 s^{-1}$ is the rate at which the two photons decay and the hydrogen ground state changes from $  2 s \rightarrow 1 s$ state, and $  K=\lambda_\alpha^3 /(8 \pi H(\it z  ))$.

The collisional ionization coefficient is \citep{minoda2017thermal}
 \begin{equation}
  \gamma_i\left(T_K\right)=0.291 \times 10^{-7} \times U^{0.39} \frac{\exp (-U)}{0.232+U} \rm cm^3 S^{-1}, 
\end{equation}
\noindent where $  U= \left|E_{1 s} / k_{B} T_K\right|$.
The rate of change of the $x_{\rm HI}$ with respect to redshift is \citep{pritchard201221}
\begin{equation}
  \frac{\it \mathrm{d}x_{\rm HI}}{\it{~\mathrm{d}} z}=[\left(1-x_{\rm HI}\right) \Lambda_{HI}-\alpha_B(T) x_{\rm HI}^2 n_{\rm H}(\it z  )]\frac{\it{\mathrm{d}} t}{\it{\mathrm{d}} z},
\end{equation}
\noindent where $1 - x_{\rm HI} \Lambda_{\rm HI}$ accounts for the ionization of HI, $\alpha_B(T) x_{\rm HI}^2 n_{\rm H}$ represents the recombination of ionized hydrogen back to the neutral state, and $\Lambda_{\mathrm{HI}}$ represents the rate of ionization per unit volume due to external processes, such as cosmic rays or UV radiation, and is the equivalent quantity in the bulk of the IGM. The balance between these processes determines the evolution of the $x_{\rm HI}$ as the Universe evolves. The evolution of $  x_{\rm HI}$ over cosmic time is influenced by various astrophysical processes, such as the formation of the first stars and galaxies, which emit ionizing radiation that can alter the state of hydrogen in the IGM. During the EoR, $x_{\rm HI}$ decreases as more regions of the IGM become ionized.

\subsection{Neutral Hydrogen Observations}\label{SEC2.3}

The hydrogen in quasars is typically more ionized than in the IGM, which increases the possibility of detecting the EoR, which is known as the Quasar Gunn-Peterson effect \citep{gunn1965density}. The Quasar Gunn-Peterson effect is observed in the spectra of quasars at high redshifts ($z \geq 6$), when the universe was still undergoing reionization. This effect is characterized by a decrease in electromagnetic emissions from the quasar at wavelengths shorter than the Ly$\alpha$ line, which is the redshift of the released light. The Gunn–Peterson technique utilizes observations of Quasi Stellar Objects (QSOs) or quasars to detect the Gunn–Peterson trough, which is a feature resulting from the absorption of light by neutral hydrogen in the IGM \citep{fan2006constraining,butcher1967gunn}. 
 
The dark gaps in the Ly$\alpha$ and Ly$\beta$ forests of quasar spectra help probe regions with little to no detectable emission or transmission, signaling a high concentration of HI that absorbs the quasar’s light \citep{schenker2014line,jin2023nearly,mcgreer2015model}. The Ly$\alpha$ forest consists of a series of absorption lines caused by clouds of neutral hydrogen gas along the line of sight to distant quasars. These lines appear near the Ly$\alpha$ transition wavelength (1215.67\,\si{\angstrom}) and trace the structure of the IGM \citep{liu2021deep,croft2002toward}. The Ly$\beta$ transition, at 1025.72\,\si{\angstrom}, produces a less forest due to its lower optical depth \citep{gnedin2022cosmic}. Extended dark gaps in these forests suggest regions with high $  x_{\rm HI}$, indicating that reionization is still ongoing. While Ly$\alpha$ transmission spikes mark fully ionized zones at redshifts above $z > 6$, some quasar sightlines show long stretches of zero flux even at redshifts below $z < 6$. These features are interpreted as remnants of neutral hydrogen “islands” that persisted into the late stages of reionization \citep{bouwens2015reionization,schenker2014line,tinker2010large}.
 
The clustering of Ly$\alpha$ Emitters (LAEs) is typically analyzed using statistical tools such as the angular correlation function, which quantifies the excess probability of finding galaxy pairs at a given separation compared to a random distribution. During the EoR, the clustering signal of LAEs is influenced by the patchy distribution of ionized bubbles in the IGM \citep{umeda2024jwst,ouchi2018systematic}. Galaxy clusters, which are among the most massive self-gravitating structures in the Universe, generally have masses ranging from $  10^{14}h^{-1}M_{\odot}$ to $  10^{15}h^{-1}M_{\odot}$ \citep{tinker2010large}. These cluster halos are associated with rare, high-density peaks in the primordial matter distribution, consistent with hierarchical structure formation models. In contrast, typical LAEs are low-mass galaxies observed at high redshifts \citep{shimasaku2006lyalpha}. They resemble dwarf galaxies and are considered potential candidates for hosting Population III stellar remnants \citep{pacucci2025exploring}.

The Ly$\alpha$ luminosity function technique entails quantifying the number density of LAEs as a function of their luminosity \citep{xu2023thesan}. This function provides insights into the star formation rate, galaxy evolution across cosmic time, and the neutral hydrogen content of the IGM during the EoR \citep{umeda2024jwst,morales2021evolution,inoue2018silverrush,ouchi2010statistics}. Significant efforts have been expended to precisely measure and characterize the observed Ly$\alpha$ luminosity function at $z \simeq  5-7$ \citep{wold2022lager,konno2018silverrush,santos2016lyalpha}. To understand the luminosity function, it is necessary to know how Ly$\alpha$ sources work and how Ly$\alpha$ photons travel through the IGM. These factors are affected by the local environment and how galaxies emit light \citep{kageura2025census,mason2019inferences}.

The Ly$\alpha$ damping wing absorption is a well-established phenomenon observed in the spectra of gamma-ray bursts (GRBs) and other high-redshift sources. Neutral hydrogen in the IGM absorbs Ly$\alpha$ photons, imprinting characteristic features in the spectra of distant GRBs and galaxies. As an instance, the spectrum of GRB 050904 at $z = 6.295$ shows a red damping wing, which lets us put limits on the neutral hydrogen fraction of $  x_{\rm HI} = 0.17^{+0.17}_{-0.17}$ \citep{totani2006implications}. GRB 130606A at $z = 5.912$ offered a substantial dataset owing to its intense optical and deep near-infrared (NIR) afterglow, facilitating extensive absorption spectroscopy and indicating a low HI column density in its host galaxy \citep{castro2013grb}. Recent JWST observations have expanded damping wing analyses to galaxies at $z = 7-12$, providing measurements of volume-averaged $x_{\rm HI}$ and ionized bubble sizes during the Epoch of Reionization \citep{umeda2024jwst}.

The Ly$\alpha$ equivalent width (EW) \citep{nakane2024lyalpha,tang2024jwst} distribution of Lyman Break Galaxies (LBGs) \citep{curtis2023spectroscopic,hsiao2023jwst} is an important factor for studying early galaxies and the EoR. LBGs are high-redshift galaxies distinguished by their intense UV continuum and a noticeable decline in flux near the Lyman limit, attributable to absorption by neutral hydrogen \citep{hoag2019constraining,mason2019inferences,whitler2020impact,tilvi2014rapid,bolan2022inferring,morishita2023early}. Observations of the Ly$\alpha$ EW distribution at $z \simeq 3$ illustrate a wide range of galaxies with different Ly$\alpha$ features, which is a sign of how their surroundings and evolutionary stages are different \citep{cooke2013nurturing,foran2023lyman}. Stronger Ly$\alpha$ emission (i.e., higher EW) tends to correlate with bluer UV colors, lower metallicities, reduced stellar masses, fainter rest-frame UV luminosities, lower star formation rates, harder ionizing spectra, and more compact morphologies. The strength of Ly$\alpha$ emission or absorption also shows sensitivity to the surrounding galactic environment. The detectability of Ly$\alpha$ in the early universe supports the prevailing galaxy formation, where more massive and luminous LBGs reside in regions of higher mass overdensity. These galaxies exhibit stronger clustering than their lower-mass, less luminous counterparts, identified as LAEs \citep{talia2023vandels}.

The near-zone regions of quasars that are located close to the background quasar are influenced by ionizing radiation from the quasar itself, nearby quasars, and the UV background. These regions exhibit distinct spectral features that enable the study of the IGM at redshifts greater than 6 through Ly$\alpha$ forest transmission \citep{bolton2011neutral}. The size of a near zone depends on the interplay between the quasar’s ionizing output and the local density of neutral hydrogen in the IGM. High-resolution cosmological simulations are used to model these near zones and the surrounding IGM, helping interpret observed spectra and constrain the properties of the EoR \citep{keating2015probing}. For example, simulations of the near zone around the quasar ULAS J1120+0641 at $z = 7.1$ suggest that the IGM was still significantly neutral at that time \citep{greig2019constraints}.
 
Absorption of Ly$\alpha$ photons by the IGM can be studied using the rest-frame UV continuum of high-redshift sources. Estimates of $  x_{\rm HI}$ at various redshifts have been derived by fitting the damping wing absorption profile across a range of $  x_{\rm HI}$ values. However, the QSOs and GRBs at redshifts beyond $z \simeq 7$ limit our ability to probe the ionization state of the IGM at these epochs \citep{matsuoka2023quasar,wang2020significantly}. Detecting the UV continuum at such high redshifts via spectroscopy is particularly challenging due to the requirement for deep NIR data. Often, the Ly$\alpha$ emission line is the only observable feature in the optical and NIR spectra of high-redshift galaxies \citep{jung2020texas}. As a result, Ly$\alpha$ is widely used as a reliable tracer of star formation activity \citep{dijkstra2010seeing,scarlata2009effect,malhotra2004desperately}. Recent observations with the JWST have demonstrated its capability to perform NIR spectroscopic measurements, enabling the detection of Ly$\alpha$ damping wing absorption and facilitating constraints on the $  x_{\rm HI}$ and ionized bubble sizes during the EoR \citep{boker2023orbit,gardner2023james,nakajima2023jwst,rieke2023performance,rigby2022characterization}.

Following the EoR, the Universe transitioned from a predominantly neutral state to an ionized one. Residual neutral hydrogen persists in dense regions such as damped Ly$\alpha$ systems (DLAs) and Lyman limit systems (LLSs), which are considered high-redshift analogs of HI-rich galaxies and serve as vital probes of cosmic evolution \citep{khandai2011detecting}. These systems trace the distribution and abundance of neutral gas across cosmic time. Radio measurements from galaxy redshift surveys facilitate the identification of the HI mass function. Probes of $  x_{\rm HI}$ at high redshifts are crucial for comprehending galaxy formation, the cosmic baryon cycle, and the early phases of the Universe. Such observations help bridge the gap between the reionization era and the present-day Universe \citep{bagla1997neutral}.

\begin{table*}[!t]
\centering
\begin{tabular}{ |c|c|c| }
 \hline
 \multicolumn{3}{|c|}{Observational Data} \\
 \hline
  Redshift(z) &  $  x_{\rm HI}$   &  References \\
 \hline
  5.03 & $0.000055_{-0.0000165}^{+0.0000142}$  & \citep{fan2006constraining}\\
 5.25 & $0.000067_{-0.0000244}^{+0.0000207}$ & \citep{fan2006constraining}\\
 5.45 & $0.000066_{-0.0000301}^{+0.0000247}$ &  \citep{fan2006constraining}\\
 5.6 &  $0.09_{-0.09}$ &  \citep{mcgreer2015model}\\
 5.65 & $0.000088_{-0.000046}^{+0.0000365}$  & \citep{fan2006constraining} \\
 5.85 & $0.00013_{-0.000049}^{+0.0000408}$  & \citep{fan2006constraining} \\
 5.9 & $0.11_{-0.11}$ & \citep{mcgreer2015model}\\
 6.1 & $0.00043_{-0.0003}^{+0.0003}$ &  \citep{fan2006constraining} \\
 6.295 & $0.17_{-0.17}$ &  \citep{totani2006implications} \\
 6.3 & $0.79_{-0.04}^{+0.04}$ &  \citep{jin2023nearly}\\
 6.5 & $0.03_{-0.03}^{+0.03}$ &  \citep{jin2023nearly}\\
 6.6 & $0.15_{-15}^{+0.15}$ &  \citep{ouchi2018systematic}\\
 6.6 & $0.08_{-0.05}^{+0.08}$ &  \citep{morales2021evolution}\\
 6.6 & $0.2_{-0.2}^{+0.2}$ &  \citep{ouchi2010statistics}\\
 $6.7_{-0.2}^{+0.2}$& $0.25_{-0.25}$ &  \citep{bolan2022inferring}\\
 6.7 & $0.94_{-0.09}^{+0.06}$ &  \citep{jin2023nearly}\\
  7 & $0.25_{-0.25}^{+0.25}$ &  \citep{ouchi2018systematic}\\
 7 & $0.28_{-0.05}^{+0.05}$ &  \citep{morales2021evolution}\\
 7 & $0.39_{-0.09}^{+0.12}$ &  \citep{schenker2014line}\\
 7 & $0.59_{-0.14}^{+0.12}$ &  \citep{whitler2020impact}\\
7.085 & $0.1^{+0.9}$ &  \citep{bolton2011neutral} \\
 7.0851 & $0.48_{-0.26}^{+0.26}$ &  \citep{davies2018quantitative} \\
 $7.1$ & $0.4_{-0.19}^{+0.21}$ &  \citep{greig2017we}\\
 7.3 & $0.83_{-0.07}^{+0.06}$ &  \citep{morales2021evolution}\\
 7.3 & $0.5_{-0.3}^{+0.1}$ &  \citep{inoue2018silverrush}\\
  7.3 & $0.55_{-0.25}^{+0.25}$ &  \citep{ouchi2018systematic}\\
  7.5 & $0.56_{-0.18}^{+0.21}$ &  \citep{banados2018800}\\
 7.5413 & $0.6_{-0.23}^{+0.2}$ &  \citep{davies2018quantitative} \\
 $7.6_{-0.6}^{+0.6}$ & $0.83_{-0.11}^{+0.08}$ &  \citep{bolan2022inferring}\\
 $7.6_{-0.6}^{+0.6}$ & $0.88_{-0.1}^{+0.05}$ & \citep{hoag2019constraining} \\
 7.66 & $0.43^{+0.57}$ &  \citep{schenker2014line} \\
 $7.9_{-0.6}^{+0.6}$ & $0.76^{+0.24}$  &  \citep{mason2019inferences}\\
 8 & $0.65^{+0.35}$ & \citep{schenker2014line}\\
 8 & $0.3^{+0.7}$ &  \citep{tilvi2014rapid}\\
  \hline
\end{tabular}
\caption{This table shows observational data regarding the $  x_{ \rm HI}$ across different redshifts. The techniques used in these data are Gunn–Peterson through QSOs \citep{fan2006constraining}, QSOs spectra dark gaps of the Ly$\alpha$ and Ly$\beta$ \citep{mcgreer2015model, jin2023nearly}, LAE clustering \citep{ouchi2018systematic}, Ly$\alpha$ luminosity function \citep{morales2021evolution,inoue2018silverrush,ouchi2010statistics}, Ly$\alpha$ damping wing absorption of
GRBs \citep{totani2006implications},  Ly$\alpha$ EW distribution of LBGs \citep{hoag2019constraining,mason2019inferences,whitler2020impact,tilvi2014rapid,bolan2022inferring}, Near-Zone Quazars \citep{bolton2011neutral}, QSOs \citep{davies2018quantitative,greig2017we,banados2018800}, and LBGs \citep{schenker2014line}.}
\label{Table:1}
\end{table*}

\begin{table*}[!t]
\hspace{10mm}
\centering
\begin{tabular}{ |c|c|c| }
 \hline
 \multicolumn{3}{|c|}{Data from JWST Observations} \\
 \hline
  Redshift(z) &  $  x_{\rm HI}$  & References \\
 \hline
 $7_{-0.5}^{+1.0}$ & $0.48_{-0.22}^{+0.15}$ &  \citep{tang2024jwst}\\
 $7.12_{-0.08}^{+0.06}$ & $0.53_{-0.47}^{+0.18}$ &  \citep{umeda2024jwst}\\
 $7.44_{-0.24}^{+0.34}$ &  $0.65_{-0.34}^{+0.27}$ &  \citep{umeda2024jwst}\\
 $7.74_{-0.34}^{+0.33}$ & $0.62_{-0.36}^{+0.15}$ &  \citep{nakane2024lyalpha} \\
 7.88 & $0.45^{+0.55}$ & \citep{morishita2023early} \\
$8.28_{-0.44}^{+0.41}$ & $0.91_{-0.22}^{+0.09}$ &  \citep{umeda2024jwst}\\
$8.81_{-0.81}^{+1.19}$ & $0.81_{-0.24}^{+0.12}$ &  \citep{tang2024jwst}\\
$9.91_{-1.15}^{+1.49}$ & $0.92_{-0.1}^{+0.08}$ &  \citep{umeda2024jwst}\\
 $10.13_{-1.52}^{+3.08}$ & $0.93_{-0.07}^{+0.04}$ &  \citep{nakane2024lyalpha} \\
 $10.603_{-0.0013}^{+0.0013}$ & $0.88_{-0.88}$ &  \citep{bruton2023universe}\\
 $10.17$ & $0.9_{}^{+0.1}$ & \citep{hsiao2023jwst} \\
 $11_{-1}^{+2.2}$ & $0.89_{-0.21}^{+0.08}$ &  \citep{tang2024jwst} \\
 $11.475_{-0.045}^{+0.045}$ & $0.78_{-0.29}^{+0.12}$ &  \citep{curtis2023spectroscopic}\\
  \hline
\end{tabular}
\caption{This table shows observational data regarding the $  x_{\rm  HI}$ across different redshifts. These $  x_{\rm HI}$ data estimates through the techniques Ly$\alpha$ EW distribution of LBGs \citep{morishita2023early,bruton2023universe}, LAE clustering analysis and Ly$\alpha$ luminosity function \citep{umeda2024jwst}, LBGs \citep{curtis2023spectroscopic,hsiao2023jwst,umeda2024jwst}, and Ly$\alpha$ EWs \citep{nakane2024lyalpha,tang2024jwst}.}
\label{Table:2}
\end{table*}

 The FDM has considerable imprints on $  \tau_{e}$ and $  x_{\rm HI}$, which can be utilized for constraints. The origin of these imprints is the small-scale suppression produced by FDM during the initial phases. This suppression postpones the development of substantial dark matter halos, thereby deferring the emergence of galaxies \citep{spina2024damping}. The UV radiation from these galaxies contributes to the reionization of the IGM, indicating that this delay may be quantified using $  x_{ \rm HI}$ and the optical depth to reionization \citep{mcgreer2015model}. 

 The EoR cannot be simply observed; instead, it must rely on theoretical models and simulations. There is not a single perfect model that precisely describes the thermal history of the EoR. The intricate interplay of radiation, gas dynamics, and cosmic structures defies simple analytical solutions. The intricate physics can be captured by simulations. They model the growth of cosmic structures, star formation, and the ionization process. By utilizing simulations, a variety of models representing different physical scenarios can be created. These models can then be compared with real observational data. Simulations make the EoR tangible. How well a model aligns with actual observations, such as $  x_{\rm HI}$ data, can be quantitatively assessed. Simulations not only validate existing theories but also predict novel phenomena.

\section{Simulations and FDM Models}\label{SEC3}

In this section, we describe the numerical tools and modeling framework used in our analysis. We begin with the cosmological Boltzmann solver \texttt{CLASS}, proceed to the semi-numerical 21-cm simulation code \texttt{21cmFAST}, and then outline the hybrid \texttt{21cmFirstCLASS} framework that integrates both. Finally, we detail the FDM parameter space explored in our study.

The Cosmic Linear Anisotropy Solving System (\texttt{CLASS}) is an advanced Boltzmann code developed to provide a more accessible and adaptable coding framework for cosmologists. \texttt{CLASS} is highly organized, easily adaptable, and systematically ensures the precision of output quantities \citep{doran2005cmbeasy}. It encompasses the following species: photons, baryons, and cold dark matter, which is required merely in synchronous gauge. Furthermore, massless neutrinos and other ultra-relativistic species, an arbitrary number of massive neutrinos and other non-cold dark matter relics, a non-perfect fluid characterized by a constant linear equation of state and sound speed, as well as a cosmological constant, are optional considerations \citep{blas2011cosmic}. Any combination of adiabatic and iso-curvature modes, with any number of randomly correlated ones, can be used as initial conditions. An amplitude, tilt, and running can be entered for each auto-correlation and cross-correlation spectrum \citep{lesgourgues2011cosmic}.

The \texttt{21cmFAST} \citep{mesinger201121cmfast} is a semi-numerical code that combines analytical and numerical approaches to efficiently simulate the cosmological 21-cm signal. These include density fluctuations, halo formation, ionization patterns, spin temperature variations, the 21-cm radiation signal, and many other astrophysical parameters. It utilizes the excursion set formalism in conjunction with perturbation theory to generate these simulations quickly and accurately \citep{murray202021cmfast}. The \texttt{21cmFAST} accurately replicates the EoR by simulating the distribution of dark matter and baryonic matter in the Universe, and subsequently including the ionization regions \citep{mertens2024retrieving,xu2017islandfast}.

The \texttt{21cmFirstCLASS} code \citep{flitter202321cmfirstclass} is a computational framework designed to simulate the 21-cm signal from the EoR and cosmic dawn while self-consistently incorporating LSS evolution. It combines the semi-numerical approach of \texttt{21cmFAST} \citep{mesinger201121cmfast} with the cosmological Boltzmann solver \texttt{CLASS} \citep{blas2011cosmic}, enabling high-precision modeling of the interplay between baryonic physics and the growth of dark matter perturbations. Its modular architecture allows independent treatment of spin temperature fluctuations, X-ray heating, and inhomogeneous recombination processes \citep{cohen2017charting,smith2022thesan}. 

\texttt{21cmFirstCLASS} has been applied to predict SKA sensitivity requirements for EoR power spectrum detection \citep{munoz2019robust}, analyze Ly$\alpha$ coupling fluctuations impact on the global 21-cm signal \citep{cohen2017charting}, and model cross-correlations between 21-cm maps and CMB polarization anisotropies \citep{smith2022thesan}. In the FDM scenario, \texttt{21cmFirstCLASS} calculates the matter transfer function utilizing \texttt{AxiCLASS} \citep{poulin2018cosmological,smith2020oscillating}, subsequently transferring it to \texttt{21cmFAST} to establish the initial conditions \citep{flitter2022closing}. 

In our analysis, we explored a discrete grid in the $(m_{\rm FDM}, f_{\rm FDM})$ parameter space, covering seven representative particle masses, $m_{\rm FDM} \in \{10^{-21},\,10^{-21.5},\,10^{-22},\,10^{-22.5},\,10^{-23},\,10^{-23.5},\,10^{-24}\}~{\rm eV}$, and five fractional contributions, $f_{\rm FDM} \in \{0.02,\,0.04,\,0.06,\,0.08,\,0.10\}$. This grid design was chosen to span the range most relevant to current astrophysical constraints while providing sufficient resolution to capture potential trends in the predicted $x_{\rm HI}$.

\section{Machine Learning Methodology}\label{SEC4}

Recent advances in machine learning have enabled the extraction of complex, non-linear patterns from high-dimensional datasets \citep{ntampaka2019role,dvorkin2022machine,alzubi2018machine}. Hardware innovations, such as in-memory computing and neuromorphic circuits, have further expanded the feasibility of these methods \citep{smagulova2019survey}. Among the most widely used approaches are neural networks, which are computational architectures inspired by the structure and function of the biological brain \citep{sazli2006brief}. In our framework, neural networks are employed to learn the mapping between simulated and observed $x_{\mathrm{HI}}$, incorporating non-Gaussian, multivariate uncertainties in both $x_{\mathrm{HI}}$ and redshift $z$. This section outlines the core architectures used in our hybrid model.

\subsection{Convolutional Neural Networks}\label{SEC4.1}

CNNs are a class of deep neural networks particularly effective for extracting spatially localized features from structured, grid-based data \citep{yamashita2018convolutional}. A typical CNN consists of an input layer, multiple convolutional and pooling layers, and one or more fully connected layers \citep{li2021survey}.

The convolutional layers apply trainable kernels (filters) to detect local patterns, progressively building hierarchical feature representations \citep{aloysius2017review}. In continuous form, the convolution of two functions $f$ and $g$ is given by
\begin{equation}
  (f * g)(t) = \int_{-\infty}^{\infty} f(\tau)\, g(t - \tau)\,\mathrm{d}\tau,
\end{equation}
illustrating how one function modulates another \citep{ajit2020review}. In practice, CNNs use discrete convolutions over pixel or voxel grids.

Pooling layers reduce spatial resolution, improving computational efficiency and feature invariance. Max pooling, for example, selects the maximum activation within a local region \citep{jie2020runpool}. This can be expressed as
\begin{align}
  x^{l}_{j} &= f^{l}(u^{l}_{j}), \\
  u^{l}_{j} &= \alpha^{l}_{j} \cdot f_{\mathrm{pooling}}^{l}(x^{l-1}_{j}) + b^{l}_{j},
\end{align}
where $x^{l}_{j}$ and $u^{l}_{j}$ are the output and net input of the $j$-th channel in layer $l$, $b^{l}_{j}$ is the bias term, $f^{l}(\cdot)$ is the activation function, $\alpha^{l}_{j}$ is the pooling weight, and $f_{\mathrm{pooling}}^{l}$ denotes the downsampling operation \citep{cong2023review}.

Fully connected layers integrate high-level features into final predictions \citep{krichen2023convolutional}, with parameters optimized via backpropagation and gradient descent \citep{gu2018recent,patil2021convolutional}. While CNNs are most effective for two-dimensional data, one-dimensional CNNs have been developed for sequential inputs \citep{kiranyaz20211d}, achieving strong performance in time-series analysis, audio processing, and galaxy morphology classification \citep{cavanagh2021morphological}. In our application, CNN layers are used to extract correlated patterns in the joint $x_{\mathrm{HI}}$–$z$ probability distributions derived from observational constraints.

\subsection{Recurrent Neural Networks}\label{SEC4.2}

RNNs are designed to process sequential data by maintaining a hidden state that evolves over time, enabling the modeling of temporal dependencies \citep{schmidhuber2015deep,yang2020lstm}. However, standard RNNs suffer from vanishing and exploding gradients \citep{pascanu2013difficulty}, computational inefficiency for long sequences \citep{prabowo2018lstm}, and difficulty retaining long-term context \citep{bengio1994learning}.

Long Short-Term Memory (LSTM) networks \citep{hochreiter1997long} address these issues using gated memory cells. The input, forget, and output gates are defined as
\begin{equation}
\begin{aligned}
i_t &= \sigma(W_i \cdot [h_{t-1}, x_t] + b_i), \\
f_t &= \sigma(W_f \cdot [h_{t-1}, x_t] + b_f), \\
o_t &= \sigma(W_o \cdot [h_{t-1}, x_t] + b_o),
\end{aligned}
\end{equation}
where $\sigma$ is the sigmoid activation, $h_{t-1}$ is the previous hidden state, and $W$, $b$ are weights and biases \citep{van2020review}. The cell state $C_t$ is updated via
\begin{equation}
\begin{aligned}
\tilde{C}_t &= \tanh(W_C \cdot [h_{t-1}, x_t] + b_C), \\
C_t &= f_t \odot C_{t-1} + i_t \odot \tilde{C}_t,
\end{aligned}
\end{equation}
and the hidden state is
\begin{equation}
h_t = o_t \odot \tanh(C_t).
\end{equation}
Gated Recurrent Units (GRUs) \citep{chung2014empirical} simplify this architecture by combining the cell and hidden states, using only a reset gate $r_t$ and update gate $z_t$
\begin{equation}
\begin{aligned}
r_t &= \sigma(W_r \cdot [h_{t-1}, x_t] + b_r), \\
z_t &= \sigma(W_z \cdot [h_{t-1}, x_t] + b_z), \\
\tilde{h}_t &= \tanh(W \cdot [r_t \odot h_{t-1}, x_t] + b), \\
h_t &= (1 - z_t) \odot h_{t-1} + z_t \odot \tilde{h}_t.
\end{aligned}
\end{equation}
Both LSTMs and GRUs mitigate vanishing gradients, with LSTMs offering finer memory control and GRUs providing faster training and fewer parameters \citep{shewalkar2019performance}. In our framework, recurrent layers are used to capture sequential dependencies in redshift evolution, enabling the network to model how $x_{\mathrm{HI}}$ changes over cosmic time under different FDM scenarios.

\subsection{Architecture and Design}\label{SEC4.3}

Hybrid CNN--RNN models have proven highly effective at capturing both spatial and sequential dependencies in complex datasets \citep{shi2019hybrid,siraj2020hybrid,halbouni2022cnn,demiss2024application}. By integrating these architectures, such models often achieve superior predictive accuracy and computational efficiency \citep{zeng2022parking}. Our neural network for FDM simulation outputs adopts a hybrid design that combines recurrent, convolutional, and deep fully connected layers. This architecture is tailored to capture spatial and temporal patterns in the uncertainties of $x_{\mathrm{HI}}$.

Spatial dependencies within each uncertainty map are modeled through CNN layers, while LSTM and GRU layers capture temporal correlations across timesteps (i.e., redshifts). Each model begins with a CNN layer, followed by a GRU layer and then an LSTM layer. This sequence allows the recurrent components to extract temporal patterns, while the convolutional and dense layers process spatial feature relationships. The design ensures that both dimensions of the data, spatial structure and temporal evolution, are jointly learned.

For optimization, we employ the Adaptive Moment Estimation (Adam) algorithm \citep{kingma2014adam} rather than classical stochastic gradient descent. Adam adaptively adjusts learning rates by combining the benefits of AdaGrad \citep{duchi2011adaptive} and RMSProp \citep{ruder2016overview}, estimating both first- and second-order moments of the gradients. The base learning rate, a critical hyperparameter controlling update magnitudes and convergence speed, is tuned via Bayesian optimization to ensure stable convergence without excessive training time. The network is trained to minimize mean squared error (MSE) loss.

To promote strong generalization and prevent overfitting, we incorporate two regularization strategies. First, dropout layers randomly deactivate 15\% of neurons in each main layer during training, reducing over-reliance on specific nodes \citep{ying2019overview}. Second, early stopping monitors performance on a randomly selected 20\% validation set each epoch, halting training when validation loss ceases to improve \citep{tripathi2024extracting}. This dual approach ensures that the model maintains predictive reliability on unseen data.

The complete system balances temporal sequence modeling through recurrent layers with spatial feature extraction in convolutional layers and feature integration in deep neural components. We assess uncertainty management via an uncertainty-weighted fine-tuning scheme, where observational errors multiplicatively scale each sample's loss before reduction. Gradients are indirectly scaled through the weighted loss, with no separate gradient manipulation. The resulting predictions are further refined through Bayesian inference, weighting residuals by their likelihood under NUTS-optimized PDFs. Residuals with high posterior probability contribute more strongly to fitness metrics, while outliers are automatically downweighted.

We incorporate observational $x_{\mathrm{HI}}$ measurements and their associated uncertainties from JWST by estimating the joint probability distribution of $x_{\mathrm{HI}}$ and redshift $z$ using Bayesian inference with the NUTS Sampler. This yields non-Gaussian, multivariate PDFs that capture correlated uncertainties, allowing high-likelihood regions in the joint parameter space to influence the model more strongly while naturally down-weighting improbable deviations.

For $x_{\mathrm{HI}}$, the NUTS-fitted posteriors provide $N_{x_{\mathrm{HI}}} = 14{,}000$ samples that preserve its empirical non-Gaussian structure. For $z$, three discrete redshift values without associated uncertainties are represented as 1-vectors, ensuring that $x_{\mathrm{HI}}$ uncertainties are fully retained in the joint construction; this produces $N_{z} = 12{,}000$ effective samples. The separable joint support is then formed via the outer product of the two marginals, resulting in 26{,}000 paired $(x_{\mathrm{HI}}, z)$ instances that densely cover the observationally allowed domain while preserving the one-dimensional distributions.

Within this framework, two complementary error measures are defined. The prediction--observation error quantifies the deviation between model outputs and JWST measurements, explicitly incorporating measurement uncertainties from the non-Gaussian, multivariate PDFs. The prediction--simulation error measures the discrepancy between predictions and the outputs of each simulated FDM scenario, reflecting the degree of theoretical alignment.

\begin{figure*}[!ht]
    \centering
    \includegraphics[width=\textwidth]{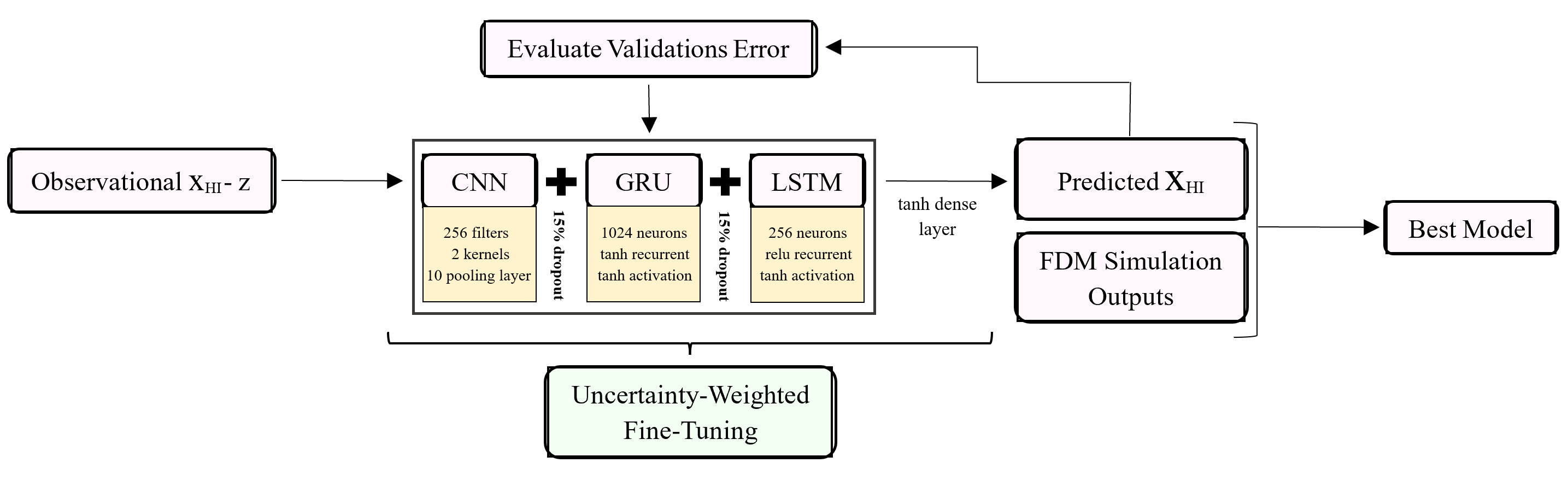}
    \caption{Schematic of the proposed hybrid CNN–RNN architecture for modeling $x_{\mathrm{HI}}$ and its uncertainties from FDM simulation outputs. The network begins with a CNN layer that extracts spatial features within each uncertainty map, followed by a GRU layer to refine temporal representations with reduced parameter complexity, and an LSTM layer to capture long-range temporal dependencies across redshift steps. Dropout layers (15\% rate) are applied to each main block to improve generalization, and early stopping is used to prevent overfitting. Uncertainty management is incorporated via an uncertainty-weighted fine-tuning scheme, where observational errors multiplicatively scale each sample’s loss before reduction, indirectly influencing gradients through the weighted loss without any separate gradient manipulation.
}
    \label{fig:schem}
\end{figure*}

Finally, the trained model evaluates competing FDM scenarios against $x_{\mathrm{HI}}$ observations, using exclusively JWST data to ensure constraints are based on the most precise current measurements. This approach enables objective discrimination between theoretical models while preserving physically consistent interpretations. The schematic of our proposed hybrid CNN--RNN architecture is shown in Fig.~\ref{fig:schem}.

\subsection{Error and Evaluation}\label{SEC4.4}

Reliable error estimation is essential for drawing statistically defensible conclusions from model predictions. To assess both predictive accuracy and generalizability, we adopt two complementary metrics \citep{shanmugasundar2021comparative}: MSE and the coefficient of determination ($R^2$). The MSE is defined as
\begin{equation}
    \mathrm{MSE} = \frac{1}{n} \sum_{i=1}^{n} \left( y_i - \hat{y}_i \right)^2,
\end{equation}
where $n$ is the number of observations, $y_i$ the measured value, and $\hat{y}_i$ the model prediction. This statistic measures the average squared deviation between predictions and targets; its quadratic form makes it particularly sensitive to large residuals, which is advantageous for detecting outliers and diagnosing overfitting \citep{hodson2021mean}.

The coefficient of determination measures the fraction of variance in the dependent variable explained by the model \citep{chicco2021coefficient}:
\begin{equation}
    R^2 = 1 - \frac{\sum_{i=1}^{n} \left( y_i - \hat{y}_i \right)^2}{\sum_{i=1}^{n} \left( y_i - \bar{y} \right)^2},
\end{equation}
where $\bar{y}$ denotes the mean of the observed data. An $R^2$ value of unity corresponds to perfect prediction, while a value of zero indicates no improvement over the mean.

Model evaluation involves two key steps: (i) assessing the ability to manage uncertainties through an uncertainty-weighted fine-tuning scheme, in which observational errors are applied as a multiplicative scaling factor to each sample's loss term before reduction. The gradient is indirectly scaled because it is computed from this weighted loss, without any separate gradient manipulation step. (ii) Testing the physical plausibility of the results against theoretical expectations across the $(m_{\mathrm{FDM}}, f_{\mathrm{FDM}})$ parameter space. In these assessments, MSE quantifies squared deviations, while $R^2$ measures explained variance.

The dataset was randomly divided into two non-overlapping subsets to support both model training and performance evaluation. A total of 80\% of the data was allocated for training, providing the model with a sufficiently large and diverse sample from which to learn underlying patterns and relationships, while the remaining 20\% was held out for validation to ensure that model performance could be assessed on unseen data and to reduce the risk of overfitting. This approach was complemented by early stopping, which monitored validation performance and halted training when no improvement was observed over a predefined number of epochs, thereby preventing unnecessary parameter updates and preserving model robustness. To further mitigate overfitting and improve generalization, dropout layers (15\%) were incorporated into the network architecture, randomly deactivating a fraction of neurons during each training iteration.

Across all metrics, the hybrid CNN--RNN architecture exhibits strong agreement with the underlying data, accurately reproducing the expected redshift evolution of $x_{\mathrm{HI}}$. The close correspondence with theoretical FDM predictions, combined with robust statistical performance, supports both the reliability and the physical plausibility of the proposed approach. The convolutional–recurrent neural network was first trained on observational data together with their associated uncertainties, during which machine--observation errors were monitored to guide learning. After training, the selected model was evaluated against the underlying simulation outputs, yielding an $R^2$ of $0.9754$ and MSE of $1.93 \times 10^{-3}$. These results indicate that the model, having learned from noisy observational inputs, was able to reproduce the simulated fields with high fidelity and minimal deviation.

\section{Results}\label{SEC5}

\begin{figure*}[!htbp]
    \centering
    \includegraphics[width=\textwidth]{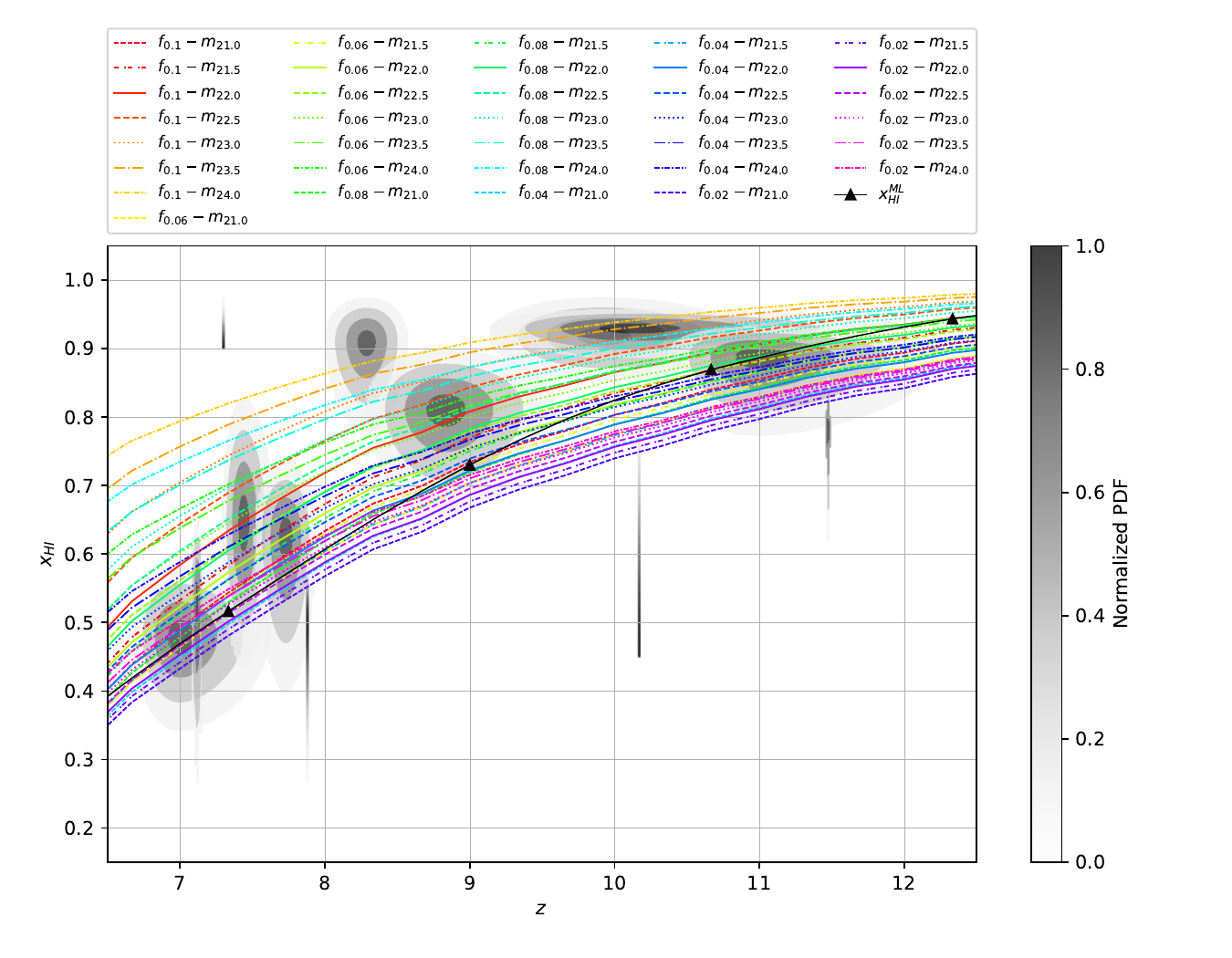}
    \caption{This figure shows the normalized posterior PDF of $  x_{\rm HI}$, from $z = 6\text{-}12$. Colored dashed curves are \texttt{21cmFirstCLASS} PDFs for a grid of FDM particle masses and fractions (legend codes denote $f_{\rm FDM}$ and   log10 m/eV). The solid black curve is the uncertainty‑aware hybrid neural network posterior mean after uncertainty‑weighted fine‑tuning to observations. Grayscale contours show credible regions from JWST‑inferred $  x_{\rm HI}$ posteriors derived with NUTS sampling.}
    \label{fig:1}
\end{figure*}

\begin{figure*}[!htbp]
    \centering
    \includegraphics[width=\textwidth]{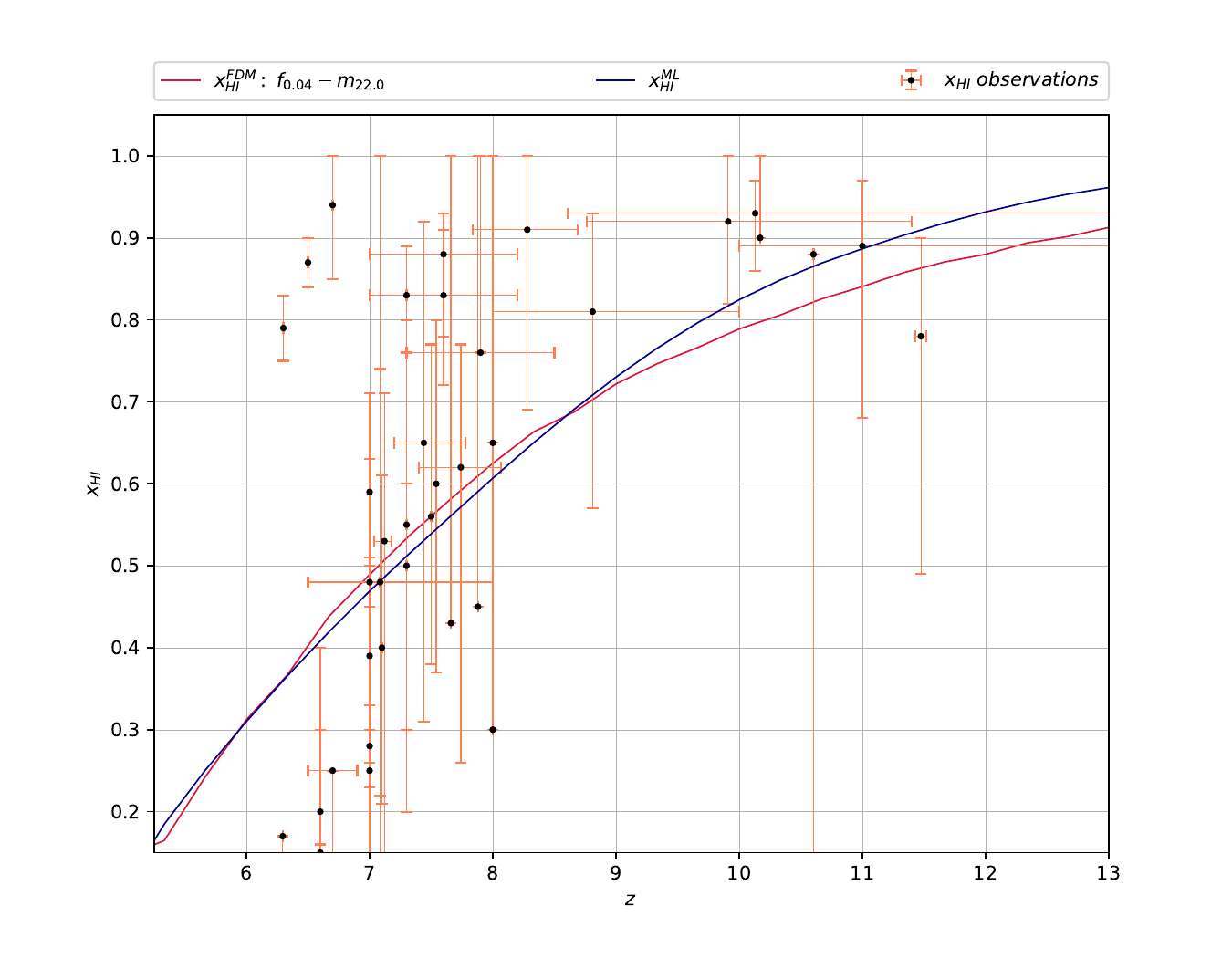}
    \caption{The FDM model displayed in this figure corresponds to the simulation (pink line) with the minimum discrepancy relative to our hybrid neural network (blue line), based on quantitative error metrics. The machine learning model, fine-tuned using JWST-inferred $  x_{\rm HI}$ posteriors, selects this configuration ($m_{\rm FDM} \simeq 10^{-22}$  eV and $f_{\rm FDM} \simeq 0.04$) with $  R^2=0.968\% $ agreement. Points with error bars compile constraints from quasar Gunn–Peterson troughs, dark-gap statistics, LAE clustering, Ly$\alpha$ luminosity functions, damping‑wing measurements, LGB Ly$\alpha$-EW inferences, and near-zone analyses (Tables \ref{Table:1}, \ref{Table:2}). The observational spread at$ z \geq 7$ highlights current tension regions; models with $  m_{\rm FDM} \simeq 10^{−22}$ eV and $f_{\rm FDM} \simeq 0.04$ at the percent level best track the central trend while maintaining consistency with upper bounds.}
    \label{fig:2}
\end{figure*}

\begin{figure*}
    \centering
    \includegraphics[width=\linewidth]{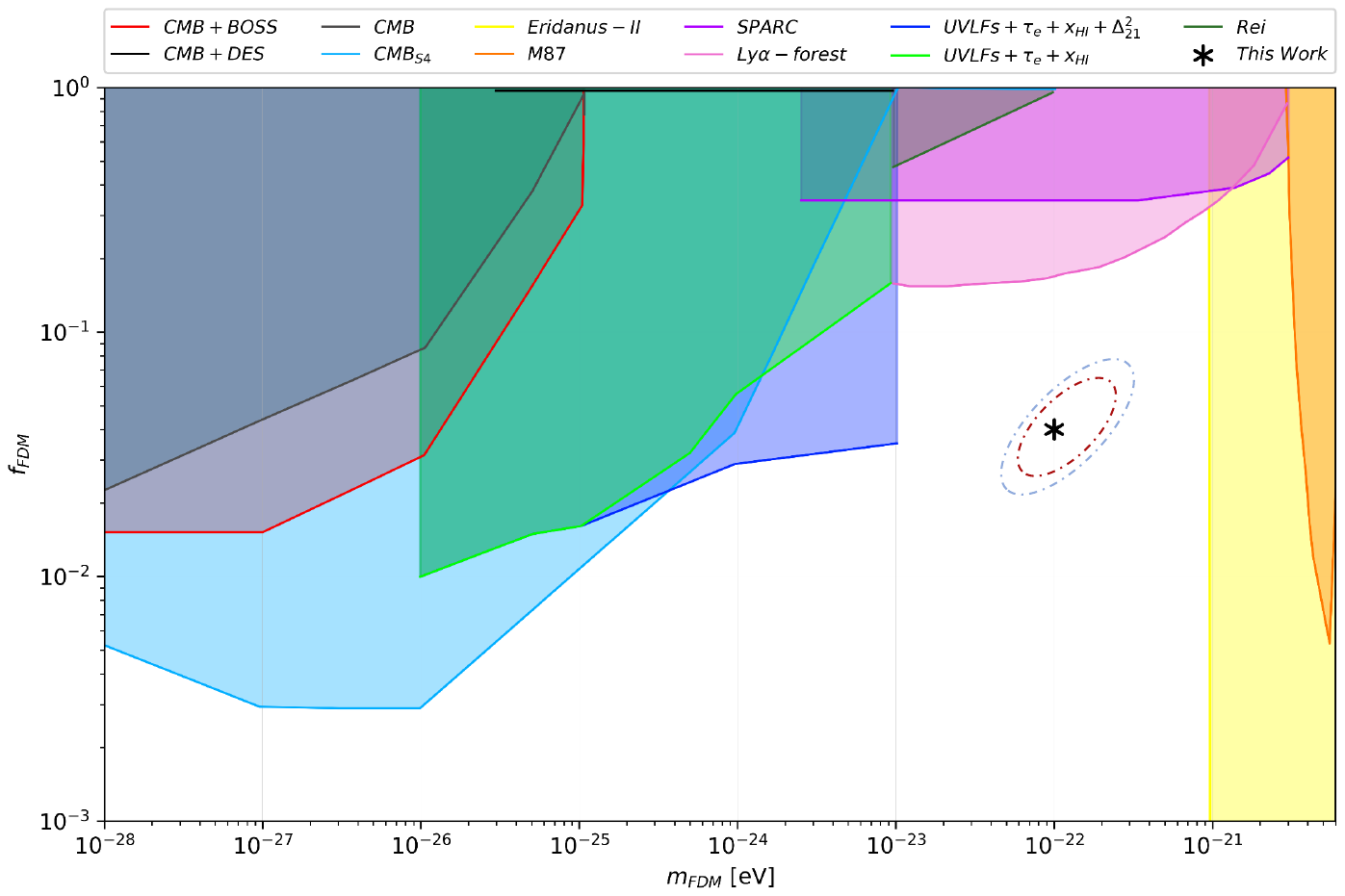}
    \caption{This figure shows parameter‑space constraints on the FDM models. Shaded and hatched regions indicate excluded masses and fractions from a range of independent probes: CMB (Planck) \citep{hlozek2015search,poulin2018cosmological}, large‑scale structure (Planck+BOSS) \citep{lague2022constraining}, Ly$\alpha$ forest \citep{irvsivc2017first,armengaud2017constraining,rogers2021strong,kobayashi2017lyman}, galaxy rotation curves (SPARC) \citep{bar2022galactic}, Dark Energy Survey \citep{dentler2022fuzzy}, M87 black hole spin \citep{tamburini2020measurement,davoudiasl2019ultralight,unal2020properties}, the half‑light radius of Eridanus‑II’s central star cluster \citep{marsh2019strong}, and reionization‑era limits from UV luminosity functions, optical depth, and $x_{\rm HI}$  \citep{lazare2024constraints}. The adopted parameter choice (“This work”) lies outside all excluded regions, satisfying the bound‑consistency criterion and avoiding the over‑suppressed structure growth and reionization delays characteristic of disfavored low‑mass models.}
    \label{fig:bounds}
\end{figure*}

\begin{figure*}[!htbp]
  \centering
  \includegraphics[width=0.45\textwidth]{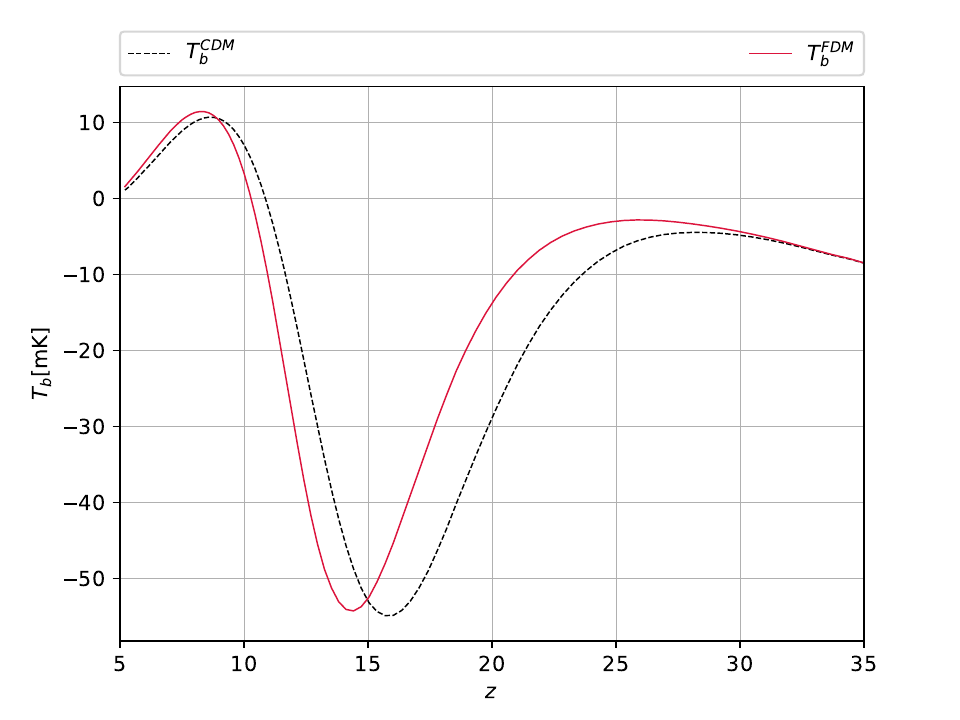}
  \includegraphics[width=0.45\textwidth]{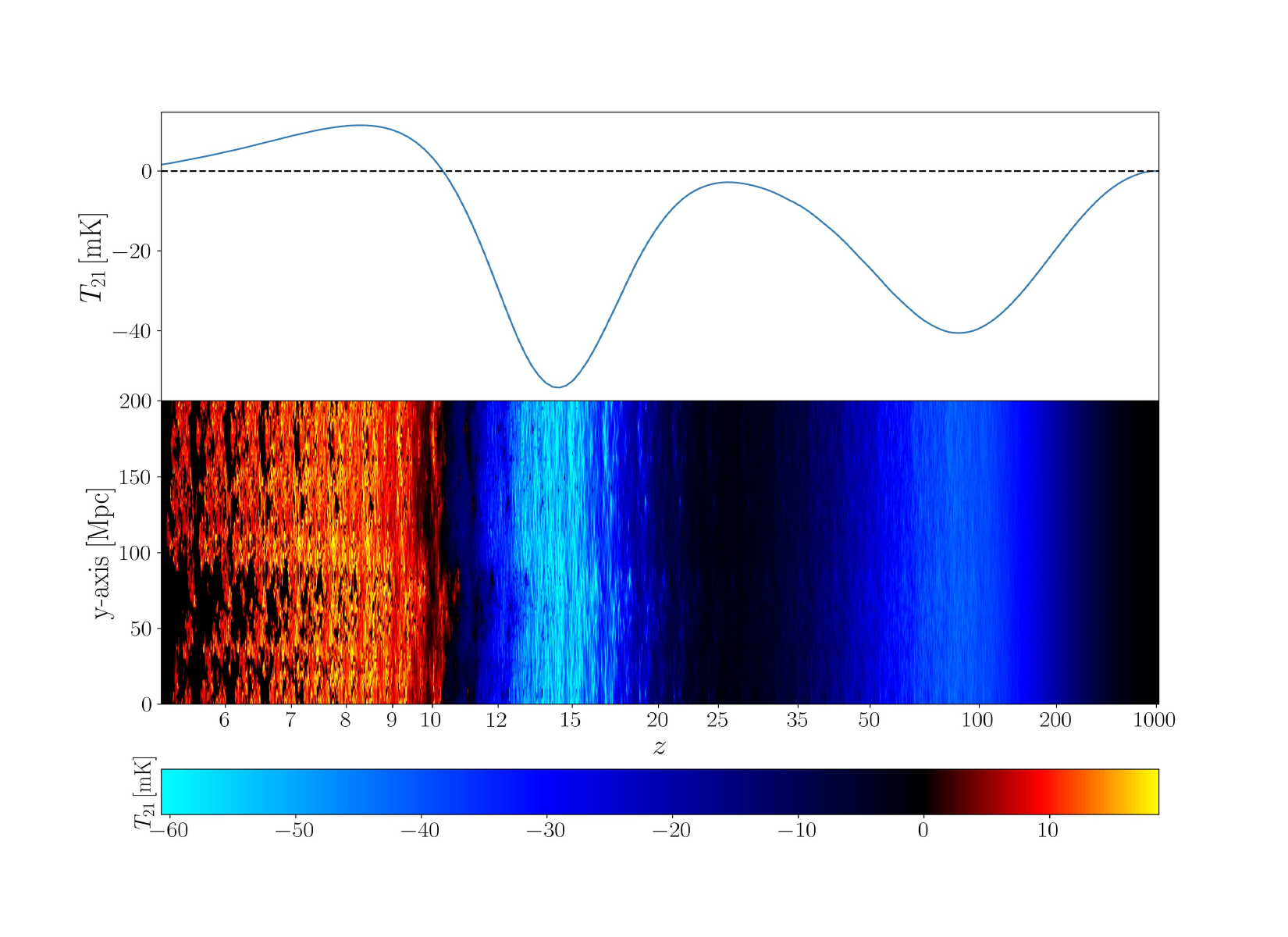}
  \caption{The left panel indicates the nominated FDM model ($m_{\rm FDM} \simeq 10^{-22} \rm eV$, $f_{\rm FDM} \simeq 0.04$) global 21-cm signal, $T_{21}(\it z)$, compares to the standard $\Lambda$CDM scenario. The FDM case shows a later onset of reionization: the first local extreme in $T_b$ occurs earlier in the FDM case, and also stays elevated for longer, consistent with reduced small-scale collapse and a slower buildup of early star formation. By contrast, the $\Lambda$CDM curve declines more steeply and reaches a sharper minimum, indicating a faster ionization history in that model setup.
 The right panel shows $  T_{21}(\it z)$ evolution and a representative light‑cone slice (line‑of‑sight redshift evolution versus transverse comoving distance) for the chosen FDM model. Blue regions (negative $T_{21}$ in mK) indicate absorption where $T_{S} < T_{\rm CMB}$, driven by Wouthuysen–Field coupling before significant X‑ray heating; red regions (positive $  T_{21}$) indicate emission where $T_{S} > T_{\rm CMB}$ due to X‑ray heating and advancing ionization fronts. }
  \label{fig:3}
\end{figure*}

We deliberately trained and tuned the hybrid model on JWST‑derived $x_{\mathrm{HI}}$ posteriors rather than on the heterogeneous dataset in Table~\ref{Table:1}. This choice rests on three methodological considerations. First, JWST data originate from a single, homogeneous instrument suite with consistent calibration, spectral coverage, and selection, minimizing cross‑survey systematics that can bias joint fits. Second, JWST analyses report full posterior distributions with asymmetric credible intervals, which we preserve in our loss function to weight residuals according to the true probability density, avoiding Gaussian or uniform‑error assumptions. Third, the strongest leverage on FDM physics lies at $z \geq 7$, where JWST spectroscopy and photometry achieve high signal‑to‑noise and reduced systematics compared to older ground‑based or pre‑JWST constraints. Restricting to these measurements reduces hidden tensions and improves parameter identifiability. Validation against the broader literature, including CMB $\tau$, Ly$\alpha$ forest limits, large‑scale structure bounds, and stellar‑dynamical constraints, confirms that the JWST‑only fit lies outside published exclusion regions and satisfies late‑epoch limits.

Fig.~\ref{fig:1} presents the uncertainty‑aware estimates of $x_{\mathrm{HI}}$ between $z = 6$ and $z = 12$. The hybrid neural network posterior mean (solid black) closely follows the envelope of \texttt{21cmFirstCLASS} simulations shown as colored dashed curves. JWST-based non-Gaussian joint likelihoods, lie largely within the model credible interval. This indicates that fine-tuning to observational inputs preserves the overall redshift evolution while avoiding overfitting to localized features.

As shown in Fig.~\ref{fig:2}, the simulation with ($m_{\mathrm{FDM}}$, $f_{\mathrm{FDM}}$) $\simeq (10^{-22}\,\mathrm{eV}, 0.04)$ exhibits the smallest weighted discrepancy with the hybrid model and achieves $R^2 = 0.975$. Observational constraints from quasar Gunn–Peterson troughs, LAE clustering, Ly$\alpha$ luminosity functions, and damping wing analyses remain consistent with this configuration. Models with similar fractions but significantly smaller masses suppress structure formation strongly, producing $x_{\mathrm{HI}}$ curves that overshoot upper bounds or diverge from the JWST-informed range at $z \geq 8$.

Fig.~\ref{fig:3} (left panel) compares the predicted global 21‑cm brightness temperature, $T_{21}(z)$, for the preferred FDM configuration ($m_{\rm FDM} \simeq 10^{-22}\,\mathrm{eV}$, $f_{\rm FDM} \simeq 0.04$) with the standard $\Lambda$CDM model. In the FDM scenario, the onset of reionization is delayed and the transition proceeds more gradually, with the first absorption trough occurring earlier and persisting for an extended period. This behaviour reflects the suppression of small‑scale structure formation, which slows the buildup of early star‑forming galaxies. By contrast, the $\Lambda$CDM curve declines more steeply and reaches a sharper minimum, indicating a faster ionization history.

The right panel of figure~\ref{fig:3} shows the corresponding light‑cone slice, illustrating the redshift evolution of $T_{21}$ along the line of sight and across transverse comoving distance. Extended blue regions (negative $T_{21}$) trace absorption phases where $T_{S} < T_{\rm CMB}$, driven by Wouthuysen–Field coupling prior to significant X‑ray heating. Red regions (positive $T_{21}$) mark emission phases where $T_{S} > T_{\rm CMB}$, associated with X‑ray heating and the advance of ionization fronts. The gradual growth of ionized bubbles and the persistence of large neutral patches in the FDM model highlight the impact of suppressed small‑scale power on the morphology and timing of reionization.

The bound consistency criterion is also satisfied as demonstrated in Fig.\ref{fig:bounds}. This includes CMB bounds from Planck \citep{hlozek2015search,poulin2018cosmological}, large scale structure from Planck combined with BOSS \citep{lague2022constraining}, Ly$\alpha$ forest limits \citep{irvsivc2017first,armengaud2017constraining,rogers2021strong,kobayashi2017lyman}, galaxy rotation curves from SPARC \citep{bar2022galactic}, Dark Energy Survey exclusion of $f_{\mathrm{FDM}} = 1$ for $10^{-25}\,\mathrm{eV} \leq m_{\mathrm{FDM}} \leq 10^{-23}\,\mathrm{eV}$ \citep{dentler2022fuzzy}, M87 black hole spin constraints from the non observation of superradiance \citep{tamburini2020measurement,davoudiasl2019ultralight,unal2020properties}, Eridanus II half light radius limits \citep{marsh2019strong}, and reionization era limits from UV luminosity functions, optical depth, and $x_{\mathrm{HI}}$ \citep{lazare2024constraints}. This parameter choice avoids the over-suppressed structure formation and over-delayed reionization characteristic of excluded low mass models, remaining consistent with all early and late epoch observational limits.

\section{Conclusion}\label{SEC6}
This work has presented an uncertainty‑aware hybrid modeling framework to explore the reionization history in the context of FDM cosmologies. By training exclusively on homogeneous JWST‑derived $x_{\mathrm{HI}}$ posteriors, we avoided the cross‑survey systematics that often complicate joint analyses and achieved a robust match to high‑redshift constraints. The selected model, characterized by $m_{\rm FDM} \simeq 10^{-22}\,\mathrm{eV}$ and $f_{\rm FDM} \simeq 0.04$, most closely reproduces the machine‑predicted reionization timeline and ionization morphology, derived from observational data with associated uncertainties, showing that its underlying physical parameters are consistent with the large‑scale structure and ionization patterns predicted by the simulations. Compared to the standard $\Lambda$CDM scenario, this model naturally delays and smooths the reionization process, producing distinctive signatures in the global 21‑cm signal and ionization morphology that can be tested with forthcoming observations.

We incorporated the full, asymmetric posterior distributions as per‑sample multiplicative loss weights applied before reduction. This indirectly scaled the gradients through the weighted loss, without any separate gradient manipulation. This approach preserved the true probability structure of the measurements and avoided Gaussian or uniform‑error assumptions. The resulting JWST‑only fit remains consistent with a broad suite of independent constraints, including CMB $\tau$, Ly$\alpha$ forest limits, large‑scale structure bounds, and stellar‑dynamical measurements, and lies outside published exclusion regions.

In this scenario, the onset of reionization is delayed and proceeds more gradually than in $\Lambda$CDM, with an extended first absorption trough in the global 21‑cm signal and a slower growth of ionized regions. Lightcone visualizations reveal persistent large neutral patches and a more gradual emergence of ionized bubbles, reflecting the suppression of small‑scale structure formation inherent to FDM. This parameter choice also satisfies all early‑ and late‑epoch bounds, avoiding the over‑suppressed structure formation and excessively delayed reionization characteristic of low‑mass models that conflict with current observational constraints..

Looking ahead, incorporating complementary probes such as CMB optical depth measurements, high‑redshift UV luminosity functions, and 21‑cm power spectrum limits into a unified, uncertainty‑aware likelihood would help tighten constraints and reduce degeneracies between dark matter and astrophysical parameters. The growing volume and diversity of high‑quality observational data present an opportunity to retrain and further calibrate the machine‑learning models, enabling them to capture a broader range of reionization histories. In addition, likelihood‑free techniques such as simulation‑based inference (SBI) could be employed to more efficiently explore the parameter space and quantify uncertainties in a fully Bayesian framework. Physics‑informed machine‑learning architectures could further improve interpretability and ensure that learned representations remain consistent with underlying physical laws. 

Further validation against larger suites of high‑resolution radiative‑transfer simulations would strengthen confidence in the emulator’s predictions, while the use of weighted morphological statistics and topology‑based measures, could provide additional discriminatory power between competing models. Finally, targeted forecasts for next‑generation facilities, including SKA, HERA, and joint JWST–Roman observations, will be essential for identifying the redshift ranges and signal features most sensitive to the physics of FDM.

\newpage
\bibliography{ref}{}
\bibliographystyle{plainnat}

\end{document}